# ANALYSING AN ANALYTICAL SOLUTION MODEL FOR SIMULTANEOUS MOBILITY

Md. Ibrahim Chowdhury[1], Mohammad Iqbal[2], Naznin Sultana[3] and Faisal Rahman[4]

[1]Department of Computer Science and Engineering, City University, Bangladesh
[2]School of Computing, Blekinge Institute of Technology, karlskrona, Sweden
[3]Department of Computer Science and Engineering, City University, Bangladesh
[4]BSc Computer Security and Forensics, University of Bedfordshire, CCNA, CEH

## ABSTRACT

*Current mobility models for simultaneous mobility have their convolution in designing simultaneous movement where mobile nodes (MNs) travel randomly from the two adjacent cells at the same time and also have their complexity in the measurement of the occurrences of simultaneous handover. Simultaneous mobility problem incurs when two of the MNs start handover approximately at the same time. As Simultaneous mobility is different for the other mobility pattern, generally occurs less number of times in real time; we analyze that a simplified simultaneous mobility model can be considered by taking only symmetric positions of MNs with random steps. In addition to that, we simulated the model using mSCTP and compare the simulation results in different scenarios with customized cell ranges. The analytical results shows that with the bigger the cell sizes, simultaneous handover with random steps occurrences become lees and for the sequential mobility (where initial positions of MNs is predetermined) with random steps, simultaneous handover is more frequent.*

## 1. INTRODUCTION

In number of researches of wireless networks, simulation is mainly used for performance evaluation. The mobility model that is used to generate the movement patterns of mobile nodes (MNs) is one of the essential simulation parameters. The mobile nodes must pose into their mobility with random step in real mobile networks. We have studied the two widely used mobility models which have commonly come under the scrutiny by the researchers. They are random walk mobility model [5] and random waypoint mobility model [4]. These models have been used in different simulation ranges and different simulation environments for location management in wireless network. As movements of MNs in random mobility models [1] [2] involve total randomness, the unrealistic simultaneous moving behaviors are invoked and could invalidate the network evaluations [3]. Thus we considered the random movement of MNs.

Simultaneous mobility is the phenomenon that occurs while both of the communicating mobile nodes (MNs) change their location simultaneously without breaking an ongoing session between them [9]. The problem arises when there are two MNs involve a communication session in normal state, and they both move such that the binding updates (assumed to do not contain information about future moves of the sending Mobile Node) that they send to each other are both lost through belated arrival. As a result the communication session never returns from interrupted to normal state and a broken connection occur [10]. One of the key issues is simultaneous mobility is the observance of simultaneous handover. In our previous research [6], we have proposed a new solution approach to solve simultaneous mobility issues in seamless handover. This 'step_length' based Simultaneous Mobility Model is capable of generating the simultaneous mobility patterns and behaviours to match with research requirements. In this paper we intend to examine the analytical solution model for simultaneous mobility by simulating the model in different scenarios.

## 2. ANALYTICAL SOLUTION MODEL

We defined a unique term called 'step_length' which is the position or state change of MNs. At each step, an MN has a specific 'step_length'. The value of 'step_length' is equal to the distance travelled by the MNs during a uniform interval time Δt. 'step_length' itself is randomly generated after every interval time Δt, and it is also uniformly distributed.

For simplicity of the model, we assumed only two mobile nodes to exist within different zones of homogeneous networks, i.e., MN_0 in Zone_0 and MN_1 in Zone_1. These zones can be worked with customized ranges.

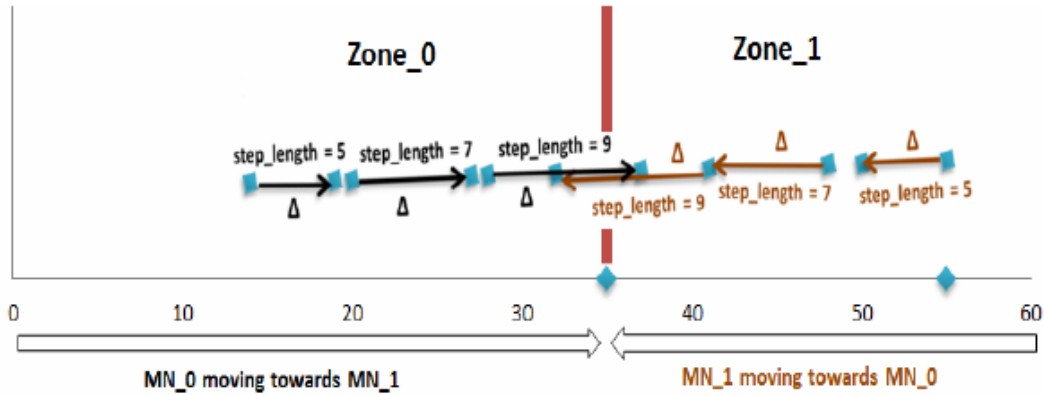

Figure 1. Simultaneous Mobility model with random 'step_length'

We further assume that two mobile nodes are simultaneously moving according to the value of 'step_length' after every interval time Δt and 'step_length' is same for both MNs for a certain move. There are no pause times between changes of direction and speed. Thus mobility of MNs can be called simultaneous. The MNs tend to start at randomly chosen position where the speed of the both MNs is uniformly chosen between specific ranges like 0 m/s to 40 m/s. If each interval time (Δt) is chosen 1 ms and random speed (v_step) is 10 m/s for a certain step, 'step_lenght' is equal to 10m (s=vt). Suppose that MN_0 and MN_1 have the symmetric movement position along x-axis. For simplicity of this model, the considerable mobility for MNs are taken by means of one dimensional values, i.e., only x axis and measured speed by means of scalar quantity (setting speed as constant). The horizontal movement takes no angle, i.e., ∡=0°. Let the values for $x_1$ is (100, 0) and $x_2$ is (150, 0). Thus the considerable movement is like that $x_1$(100, 0) to $x_2$(150, 0). That means MN is moving from the position 100 to 150.

Also 'step_length' determines the relationship between MNs past and future movement according to our design concept. The simple algorithmic formula for MNs positions is as follows:

MN_0_init.x + step_length = MN_0_new (i)

MN_1_init.x - step_length = MN_1_new (ii)

Where, x is any random value along x-axis

By this formula, we can easily implement the simultaneous mobility behaviour depicted in Figure 1. At any Δt time, MN_0 is moving towards MN_1 from Zone_0 and MN_1 is also moving towards MN_0 from Zone_1. Consequently both MNs are moving closer to each other. In the next movement, new random value of 'step_length' will be added with the previous MN_0's position (equation (i)). Similarly the same 'step_length' is deducted from MN_1's old position to detect MN_1's new location (equation (ii)). Hence, from the Table 1, we can interpret the simultaneous movements of MNs. For instance, at any random 'step_length' value '5', MN_0 is moving from the position 14 to 19 and MN_1 is also moving from 55 to 50 simultaneously at time Δt.

Table 1: Sample random values of 'step_length' and simultaneous movements of MNs

| Step_Length | MN_0 moves toward MN_1 | | MN_1 moves toward MN_0 | |
|---|---|---|---|---|
| 5 | (14,0) | (19,0) | (55,0) | (50,0) |
| 7 | (20,0) | (27,0) | (48,0) | (41,0) |
| 9 | (28,0) | (37,0) | (41,0) | (32,0) |

In Figure 1, the simulation model of simultaneous mobility is illustrated. With the randomly generated values 'step_length', MN_0 and MN_1 are simultaneously moving towards each other and when they crossed over into Zone_1 and Zone_0 respectively at the same time, simultaneous handover happens. We measure the simultaneous mobility incidents of both MNs. This occurrence plays an important role in measuring the performance of the solution. In the implementation chapter, we included all our assumptions and parameters aimed at followed model.

## 3. SIMULATION AND IMPLEMENTATION

For this model, we assume that MN_0 initially starts from a position which is located at inferior distance from MN_1's initial place. According to the architecture of this model the 'step_length' is limited within a certain range of random values, which has to be lower than ranges of Zone_0 and Zone_1. Otherwise MN's random steps may exceed the boundary of its own region. The Figure 2 demonstrates simultaneous mobility pattern with basic handover scenario. Here we can imagine an 'Overlapping Coverage' in between two adjacent cells. This 'Overlapping Coverage' is essential in measuring the handover occurrences. A handover or simultaneous handover of MNs is triggered by any random 'step_length' closer to this 'Overlapping Coverage'. The decision, when to switch from the old zone to the new one; is based on defining the minimal 'Overlapping Coverage' between two adjacent cells and the choice of random 'step_length' values by MNs. These factors for handover process are also significant in setting up an alternate solution of simultaneous mobility.

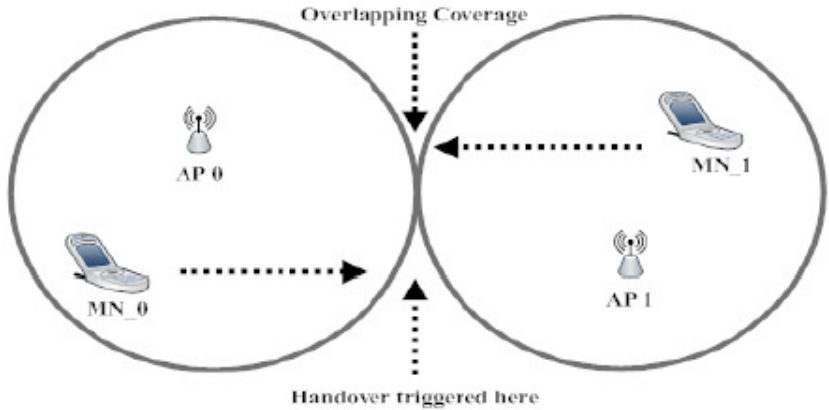

Figure 2. The Simultaneous Mobility of MN_0 and MN_1 with basic handover Scenario

We limit our solution by setting up only these two measurements amongst others factors for gaining seamless handover like user's mobility at high speed (setting speed as constant).
In this solution, simultaneous mobility is successfully observed when simultaneous handover occurs. We successfully implement and observed this very behavior by in NS2 [11] with mSCTP [8].

### 3.1 Simulations

The following sub-section includes the definitions of the performance matrices which are required to be mentioned before building the simulation scenarios. Then, we proceed with the scenarios based on our assumptions and observed experiments. This segment contains the incorporation and validation of our proposed model as a solution for simultaneous mobility problem.

#### 3.1.1 Definition of performance matrices for simulation scenarios

There are certain performance matrices involve in the simulation scenarios based on our assumptions and experiments. We define these as performance matrices which act as preconditions for understanding the simulation behaviors. The following are short description of performance metrics:

In Figure 3, we define 'Brink plane' as the minimal 'overlapping Coverage' in between Zone_0 and Zone_1, where MNs are aware of possible handover. Particularly we make the following important assumptions for handover of MNs in the overlapping region i.e., 'Brink plane':

a. The overlapping region between different subnets i.e., zones is enough large, and the sojourn time of MN in this region is larger than the time taken to perform the Add-IP operation of mSCTP [7].
b. Any MN can use two IP addresses in the overlapping region as both of the MNs are multi-homed SCTP endpoints [12].

We further assume for the forthcoming simulation scenarios that, when any of the simultaneously moving MNs (e.g., MN_0 or MN_1) touches or exceeds this 'Brink plane' a logical handover occurs. Concurrently while both of the MNs touch or crossed the 'Brink plane' at the same time of simulation, simultaneous handover will occur.

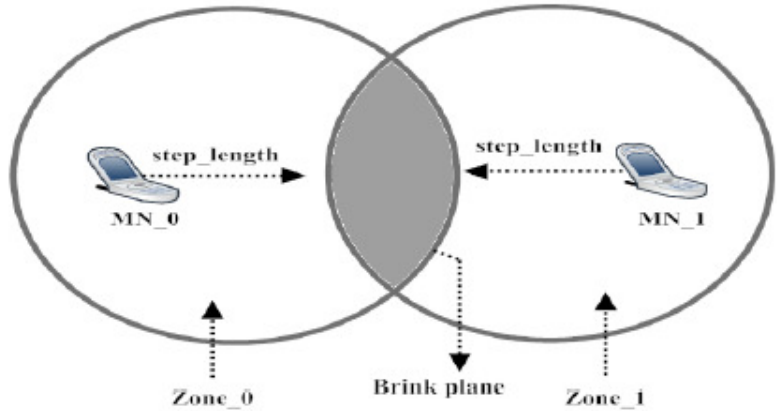

Figure 3: The simultaneous mobility scenario with 'Brink plane'

☺ Overlap: An overlap occurs when any mobile node i.e., MN_0 or MN_1 passed over their boundary zone at any certain time.
☺ MN_0 overlaps: When only MN_0 passed over the boundary of Zone_0.
☺ MN_1 overlaps: When only MN_1 passed over the boundary of Zone_1.
☺ Simultaneous Overlaps: It is the phenomenon, when both of the MNs overlap into each other's zone.
☺ No overlaps: It occurs when any of the mobile nodes from any zone does not pass their zone border. It means MN_0 or MN_1 is still moving simultaneously in their belonged areas.
☺ Sequential Handover: A sequential handover occurs while MN_0 or MN_1 or both moves step by step along with their paths in consistent time intervals and passed the 'Brink plane'.
☺ Simultaneous Handover: A simultaneous handover occurs when both MN_0 and MN_1 moving simultaneously and at the same time instance they crossed over each other's zones.
☺ MN_0 handover: It is the total number of simultaneous handover in addition with MN_0's overlapping number.
☺ MN_1 handover: It is the total number of simultaneous handover in addition with MN_1's overlapping number.
☺ Avg. 'step_length': Average 'step_length' is the summation of 'step_length' values divided by total number of simulation run for a particular simulation.

### 3.1.2. Simulation scenarios, results and observations

Different simulation scenarios are considered to measure and analyze the integrated solution approach. For this section, we propose three different scenarios for observing simultaneous mobility phenomenon and make the following assumptions for each of the scenarios:

☺ Simulation time for each run = 3600 seconds.
☺ Total Simulation = 30 times for each revisited values i.e., $V_i = V_1, V_2, \ldots, V_{30}$ where $V_n = \Sigma V_i/i$; number of sample, n = 1, 2,......., 30 and i = 1, 2,........, 30.
☺ Total time = 3600*30 = 108000 seconds = 1800 minutes = 30 hour for each sample run.

The following discussions are based on different simulation scenarios and prospective outcomes:

### 3.1.2.1 Scenerio-1: Randomly moving mobile nodes for bigger range

There are different ranges for both zones. Range for Zone_0 is from 0 to 374 units and range forZone_1 is from 376 to 750 units. The 'Brink plane' value is considered at 375 unit point.

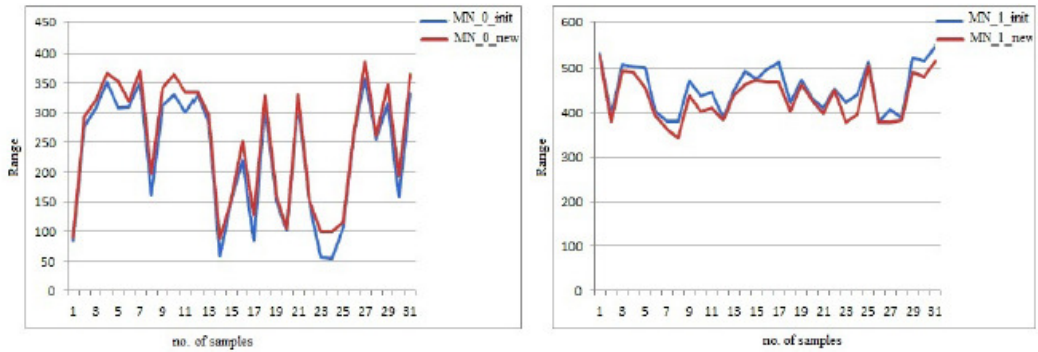

Figure 4. Showing random movements of MN_0 and MN_1 proceeding with random 'step_length'(0-50)

At each unit time Δt (1 ms), any MN is supposed to move with a random 'step_length'. We assumed in the whole simulation that this 'Brink plane' value is the point after which MN_0 or MN_1 switch over to Zone_1 or Zone_0 successively. Thus, this point is considered as the 'Brink plane' point for simultaneous handover.

From the Figure 4, we can observe that the simultaneous mobility pattern of MN_0 and MN_1 are random and non-sequential. Within the same range, but two different zones MN_0 and MN_1 are simultaneously moving.

We have built a Tcl script [8] which can initiate the random movements of mobile nodes simultaneously with random 'step_length' values. In each run of simulations there are different random movements of MN_0 and MN_1 are observed.

At every simulation run, the mobile nodes move according to generated 'step_length' value. The values of every step motion are uniformly distributed random values within 0 to 50. So, we always find a unique value of 'step_length' within this range (Table 3). In Table 2, we can observe that MNs are following the presented solution model (section 2) for simultaneous mobility. At each step, randomly generated 'step_length' value is added to the MNs initial positions (MN_0_Init or MN_1_Init) to retrieve MNs new position (MN_0_New or MN_1_New) for location update in simultaneous mobility. Following are some results generated for this specific scenario:

Table 2: Simulation results showing overlapping number

| MN_0 overlaps | MN_0 handover | MN_1 overlaps | MN_1 handover | Simultaneous overlap | No overlap | Simultaneous Handover |
|---|---|---|---|---|---|---|
| 01 time | 01 time | 01 time | 01 time | 0 time | 28 times | 0 time |

Here, with this observation, we derive the formula for calculating average 'step_length'.

$$\text{Average 'step\_length'} = \frac{\text{Summation of 'step\_length' values}}{\text{Number of simulation runs}} \quad \text{(iii)}$$

$$= 638.1 \div 30 = 21.27 \text{ approx.}: 22 \text{ steps.}$$

**Zone_0** is ranged within 0 to 374, i.e., 374 unit distances. Thus, mathematically MN_0 should cross over to Zone_1 after every 17 steps (i.e., 3374/22) to handover.

**Zone_1** is ranged within 376 to 750, i.e., 374 unit distances. Thus, mathematically MN_1 should cross over to Zone_0 after every 17 steps (i.e., 3374/22) to handover.

It takes 17+17, i.e., 34 estimated runs mathematically calculated steps to crossover two times (i.e., 1 for MN_0 overlap and 1 for MN_1 overlap) to each other's zones. Simulation is taken for 30 runs. So, mobile nodes crossing over 'Brink plane' only 2 times i.e., 34/17 times within 30 runs which statistically satisfied with estimated data.

Table 3. Data sets from randomly generated moving values of MN_0 and MN_1 for bigger range for scenario-1:

| Step_Length | Zone 0 | | Zone 1 | |
| --- | --- | --- | --- | --- |
| | MN_0_Init | MN_0_New | MN_1_Init | MN_1_New |
| 6 | 84 | 90 | 534 | 528 |
| 16 | 276 | 292 | 396 | 380 |
| 14 | 308 | 322 | 508 | 494 |
| 14 | 352 | 366 | 504 | 490 |
| 45 | 308 | 353 | 501 | 456 |
| 9 | 310 | 319 | 402 | 393 |
| 18 | 352 | 370 | 381 | 363 |
| 37 | 161 | 198 | 381 | 344 |
| 31 | 311 | 342 | 470 | 439 |
| 34 | 331 | 365 | 438 | 404 |
| 35 | 300 | 335 | 446 | 411 |
| 5 | 330 | 335 | 387 | 382 |
| 12 | 284 | 296 | 449 | 437 |
| 30 | 58 | 88 | 494 | 464 |
| 3 | 150 | 153 | 476 | 473 |
| 31 | 220 | 251 | 498 | 467 |
| 44 | 84 | 128 | 513 | 469 |
| 20 | 308 | 328 | 423 | 403 |
| 10 | 150 | 160 | 473 | 463 |
| 3 | 102 | 105 | 432 | 429 |
| 14 | 316 | 330 | 411 | 397 |
| 4 | 149 | 153 | 454 | 450 |
| 43 | 56 | 99 | 422 | 379 |
| 45 | 54 | 99 | 440 | 395 |
| 8 | 108 | 116 | 513 | 505 |
| 2 | 262 | 264 | 379 | 377 |
| 28 | 356 | 384 | 407 | 379 |
| 6 | 255 | 261 | 388 | 382 |
| 33 | 315 | 348 | 524 | 491 |
| 35 | 158 | 193 | 515 | 480 |
| 32 | 332 | 364 | 548 | 516 |

**3.1.2.2 Scenario-2: Randomly moving mobile nodes using lower range**

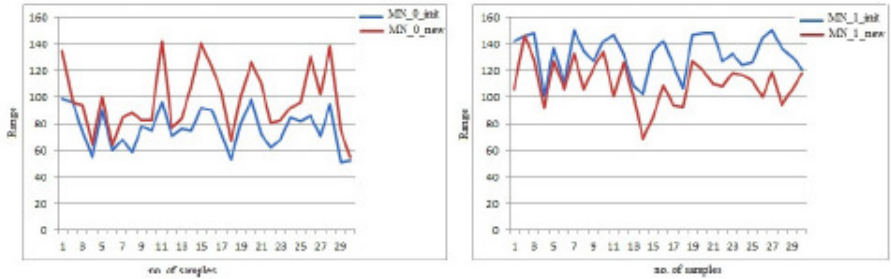

Figure 5. Showing random movements of MN_0 and MN_1 proceeding with random 'step_length' (0-50)

For analysing different viewpoints of simulation, we set the following ranges of simultaneous mobility:

Range for MN_0 is between 50 to 99 units = Zone_0
Range for MN_1 is between 101 to 150 units = Zone_1

'Brink plane' position: 100 unit point

Here, the maximum 'step_length' value is limited within 50 as before. At each unit interval time Δt, any mobile node is moving with a random 'step_length' (Table 5). In Table 5, we can observe that MNs are following the presented solution model (section 2) for simultaneous mobility. At each step, randomly generated 'step_length' value is added to the MNs initial positions (MN_0_Init or MN_1_Init) to retrieve MNs new position (MN_0_New or MN_1_New) for location update in simultaneous mobility.

For this specific scenario, we make variation only in the ranges of MNs zone and 'Brink plane'. Every parameter remained same as previous scenario. Thus, we observe the random and non-sequential mobility patters of MN_0 and MN_1 which shown in Figure 5.

Following are some results generated for this specific scenario:

Table 4: Simulation results showing overlapping number

| MN_0 overlaps | MN_0 handover | MN_1 overlaps | MN_1 handover | Simultaneous overlap | No overlap | Simultaneous Handover |
|---|---|---|---|---|---|---|
| 08 times | 13 times | 02 time | 07 times | 05 times | 5 times | 05 times |

Here, from equation (iii), the average 'step_length' comes approximately 22.

**Zone_0** is spread over 50 to 99 unit distances. Thus, MN_0 mathematically should cross over Zone_0 after every (49/21.5) = approx. 2.27 steps.
**Zone_1** is spread over 101 to 150 unit distances. Thus, MN_1 mathematically should cross over Zone_1 after every (49/21.5) = approx. 2.27 steps.

The number of simulation taken is 30 runs for one sample. So, the estimated times occurring overlap is 13.21, i.e., (30/2.27). Simulation result shows (Table 4) that total 15 times (MN_0 overlaps + MN_1 overlaps + Simultaneous overlaps) overlapping occur for MN_0 and MN_1. Thus approximately it has similarities with statistical data.

Table 5. Data sets for randomly generated MN_1 and MN_0 for lower range in scenario-2

| Step Length | Zone 0 | | Zone 1 | |
| --- | --- | --- | --- | --- |
| | MN_0 Init | MN_0 New | MN_1 Init | MN_1 New |
| 36 | 99 | 135 | 142 | 106 |
| 0 | 96 | 96 | 146 | 146 |
| 20 | 74 | 94 | 148 | 128 |
| 9 | 55 | 64 | 101 | 92 |
| 10 | 90 | 100 | 137 | 127 |
| 4 | 60 | 64 | 110 | 106 |
| 17 | 68 | 85 | 150 | 133 |
| 29 | 59 | 88 | 135 | 106 |
| 5 | 78 | 83 | 127 | 122 |
| 8 | 75 | 83 | 142 | 134 |
| 46 | 96 | 142 | 147 | 101 |
| 6 | 71 | 77 | 132 | 126 |
| 8 | 76 | 84 | 109 | 101 |
| 33 | 75 | 108 | 102 | 69 |
| 49 | 92 | 141 | 134 | 85 |
| 33 | 90 | 123 | 142 | 109 |
| 31 | 73 | 104 | 125 | 94 |
| 14 | 53 | 67 | 107 | 93 |
| 20 | 81 | 101 | 147 | 127 |
| 28 | 98 | 126 | 148 | 120 |
| 38 | 72 | 110 | 148 | 110 |
| 19 | 62 | 81 | 127 | 108 |
| 15 | 68 | 83 | 133 | 118 |
| 7 | 85 | 92 | 124 | 117 |
| 14 | 82 | 96 | 126 | 112 |
| 45 | 86 | 131 | 145 | 100 |
| 31 | 71 | 102 | 150 | 119 |
| 43 | 95 | 138 | 137 | 94 |
| 24 | 51 | 75 | 130 | 106 |
| 3 | 52 | 55 | 121 | 118 |

**3.1.2.3 Scenario-3: Sequential moving with Random 'Step_length' and Handover**

Here we assume that the mobile nodes are moving sequentially with random 'step_length' values at each run. Sequentially in a sense that the MNs initial positions are predetermined for the simulation purpose but each steps of the MNs remains random i.e., the values of 'step_length' are randomly generated. The simultaneous mobility of mobile nodes (MN_0 and MN_1) is assumed to be successive from both zones (Zone_0 and Zone_1). We consider new ranges:

Zone_0 for MN_0: 0 to 249 units
Zone_1 for MN_1: 251 to 500 units
'Brink plane' position: 250 unit point

In this scenario, it is assumed that MN_0 starts from initial position (10, 0) with random 'step_length' in Zone_0. At the same time from Zone_1, MN_1 starts to move from (500, 0) position with same random 'step_length'. At STEP-1 (Figure 6), MN_1 is moving from position (500, 0) to position (472, 0) with random 'step_length' value 28. With the same 'step_length' value, i.e., 28, MN_0 is moving from the position (10, 0) to position (38, 0) simultaneously. The detail MNs position values along with associated 'step_length' values are shown on Figure 7 and Table 6.

```
STEP-1
M 0.00100 1 (500.00, 00.00), (472.00, 00.00), 28.00
M 0.00100 0 (10.00, 00.00), (38.00, 00.00), 28.00
STEP-2
M 0.00100 1 (472.00, 00.00), (429.00, 00.00), 43.00
M 0.00100 0 (38.00, 00.00), (81.00, 00.00), 43.00
STEP-3
M 0.00100 1 (429.00, 00.00), (392.00, 00.00), 37.00
M 0.00100 0 (81.00, 00.00), (118.00, 00.00), 37.00
............................................
............................................
............................................
STEP-11
M 0.00100 1 (263.00, 00.00), (221.00, 00.00), 42.00
M 0.00100 0 (247.00, 00.00), (289.00, 00.00), 42.00
............................................
simultaneously Handover Occurs Here
```

Figure 6. MN_0 and MN_1 simultaneously moving and handovers occur at step-11

In this particular scenario; for simulation, we have only initialized the motions (random and sequential) of MN_0 and MN_1 that are simultaneously moving towards each other. Every next movement from the previous positions are generated randomly. At each run of simulation, we retune the MNs movement positions only. The 'step_length' values and MNs next movement positions are generated by the integrated Tcl code [8]. In Table 6, we can observe that MNs are following the presented solution model (section 2) for simultaneous mobility. At each step, randomly generated 'step_length' value is added to the MNs initial positions (MN_0_Init or MN_1_Init) to retrieve MNs new position (MN_0_New or MN_1_New) for location update in simultaneous mobility.

The following is one of the sample simulation results of the movement patterns from trace file [8] by:

**Formation of output**

- M-movement
- 0.001-time for each movement unit (second)
- 1/0-id of mobile node
- (_.00, _.00, 0.00),-initial position($x_1, y_1, z_1$)
- (_.00, _.00, 0.00),-new position ($x_2, y_2, z_2$)
- _.00-step_length:distance travelled in each step

In this scenario, every movement of each MN is sequential and mobility pattern is simultaneous. At every Δt (1ms) time, random 'step_length' value is generated and mobile nodes are coming nearer to each other. Simulation for this scenario continues both of the mobile nodes i.e., MN_0 and MN_1 handovers to each other's zone i.e., simultaneous handover occurs.

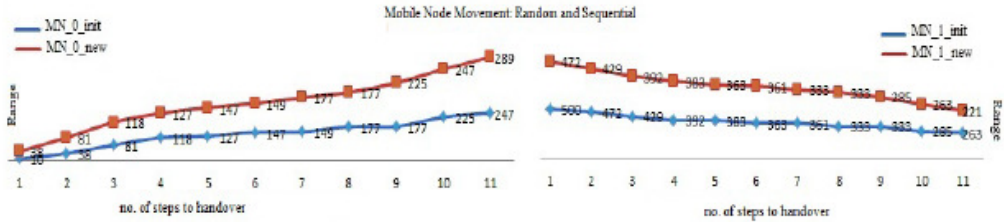

Figure 7. MN_1 and MN_0, handovers to each other's region simultaneously with sequential mobility

Table 6: Data set from a random simulation run out of 30 times, (MN_1 and MN_0, handovers to each other's region simultaneously)

| Step_Length | Zone 0 | | Zone 1 | |
| --- | --- | --- | --- | --- |
| | MN_0_Init | MN_0_New | MN_1_Init | MN_1_New |
| 28 | 10 | 38 | 500 | 472 |
| 43 | 38 | 81 | 472 | 429 |
| 37 | 81 | 118 | 429 | 392 |
| 9 | 118 | 127 | 392 | 383 |
| 20 | 127 | 147 | 383 | 363 |
| 2 | 147 | 149 | 363 | 361 |
| 28 | 149 | 177 | 361 | 333 |
| 0 | 177 | 177 | 333 | 333 |
| 48 | 177 | 225 | 333 | 282 |
| 22 | 225 | 247 | 285 | 263 |
| 42 | 247 | 289 | 263 | 221 |

Following are some results generated for this specific scenario and parameters:

Table 7. Simulation results showing overlapping number

| MN_0 overlaps | MN_0 handover | MN_1 overlaps | MN_1 handover | Simultaneous overlap | No overlap | Simultaneous Handover |
| --- | --- | --- | --- | --- | --- | --- |
| 07 times | 27 times | 03 times | 23 times | 20 times | 20 times | 20 times |

**Zone_0** is ranged within 0 to 249 units of distance. Thus, mathematically MN_0 should cross over Zone_0 after every (249/22 =) 11.31, approx. 11 steps.

**Zone_1** is ranged from 251 to 500 units of distance. Thus mathematically MN_1 should cross over Zone_1 after every (249/22 =) 11.31, approx. 11 steps.

Here, from equation (iii), we found the mean 'step_length' value is approximately 22. It is statistically measured that, out of 30 different simulations run that it would take average 11 steps every time to occur sequential handover.

Simulation results show that it would take approximately 11 steps every time for MN_0 or MN_1 or both in simultaneous handover. Thus the estimated result is closely fit to the actual result. In Table 7, we can observe that in sequential simultaneous movements of MNs, the simultaneous handover has occurred 20 times out of 30 times overlapping. So we can see that with the mean step value 11 and average 'step_length' value 22, the rate for sequential handover has increased.

**3.2 Simulation Insight**

FTP traffic is launch for mSCTP implementation in NS2 [8]. FTP is attached with SCTP agent. Simulation is started at 0.000001s. After that at 0.5s FTP traffic is generated. Until the simulation time stops which is set as 3600s, the simulation will continue for each sample. FTP needs to stop before simulation ends. In the simulation, FTP usually set-off around 20s before simulation ends. Typically at 125.001s, we initiated to call the ASCONF [7] for mSCTP. Drop-tail queue is followed to store the packets. The duplex links are set with delay of 200ms and bandwidth of 0.5Mb for interfaces used within SCTP agent. We started to record of all our simulation results after system gets stable. No results were recorded but observed during warm-up periods. Approximately 30 min warm-up time taken to ensure that the system is in steady state. Buffer is flushed after that to get more free memory utilization. We measured all the simulation results for 30 different values where each value consists of 30 different simulations. All the data sets are assumed to follow normal distribution.

## 4. CONCLUSIONS

We have analyzed our simulation work from statistical as well as technical viewpoints. Different simulations have been done with respect to different scenarios. No simultaneous handover is observed for random patterns of MN_1 and MN_0 together in Scenerio-1. The reason might be the bigger cell ranges of the MNs. Hence, we predict that lower ranges of the MNs zones may increase the probability of simultaneous handover. In Scenerio-2, both simultaneous mobility and simultaneous handover for non-sequential (random) patterns of MNs have been observed. The measurement shows that if the ranges of MNs are decreased, the number of times occurring simultaneous handover increases. We can also realize the fact that in simultaneous mobility, the number of simultaneous handover is less than the number of handover for MN_0 or MN_1 independently (Table 4). In Scenerio-3, sequential and simultaneous movement patterns of both MN_0 and MN_1 have been tested. The obtained results illustrate that the simultaneous handover rate is less than the handover rate for either MN_0 handover or MN_1 handover (Table 7).

By justifying several simulation scenarios of simultaneous and random movements of MNs, we worked on the simultaneous with sequential pattern of mobility to generate the actual nature of simultaneous mobility. The results mentioned articulate the performance of the 'step_length' based simultaneous model. It is derived that with the average 'step_length' and average steps of simultaneous movements of MNs, the estimated and concrete results were approximately the same. In future research, the movement of mobile nodes can be considered multi-dimensional instead of one dimensional to design more realistic simultaneous mobility model. Also there is scope to improve this model for large number of MNs and wide ranges of cells.

**Authors**

**Md. Ibrahim Chowdhury** pursued his MSc in Telecommunications from Blekinge Institute of Technology (BTH), Sweden in 2011. Since then, he has been working as a Lecturer in the Department of Computer Science and Engineering at City University, Bangladesh. He completed B.Sc. in Computer Science and Engineering from International Islamic University Chittgong (IIUC), Bangladesh in 2007. He also conducts IT courses in the Department of Business Administration at IIUC as an adjunct faculty. Lately he has participated in the 4th International Conference on Computational Intelligence, Modelling & Simulation in Malaysia. His present research interests include: Mobility Protocols, Simulation and Modelling, Pervasive Computing, Wireless Ad hoc Networks and Neural Networks. He is a member of IEEE and IEEE Com Soc. 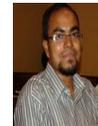

**Naznin Sultana** received B.Sc degree in Electronics and Computer Science from Jahangirnagar University, Bangladesh in 2000 (exam of 1997) & M.Sc. degree in Computer Science & Engineering from the same institution in 2010. She is a research student (Ph.D) in the department of Computer Science & Engineering, Jahangirnagar University, Bangladesh. She has teaching experience of more than 12 years in different universities such as Comilla University, Daffodil Institute of Information Technology and City University in Bangladesh. She also served as a Software Project Coordinator in Millineum Information Solutions, Bangladesh. Now she is working as an Assistant Professor in the department of Computer Science & Engineering at City University, Bangladesh. She actively participated number of international conferences. She has conference and journal papers both in national and international journals. Her research interests include Computer Vision and Pattern Recognition, Bioimaging and Image Analysis, Computer Security, Wireless communications, E-health care etc. She is an Associate member of Bangladesh Computer Council (BCC). 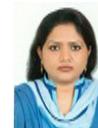

**Mr. Faisal Rahman** is a Lecturer at International Islamic University of Chittagong, a trainer of CCNA and CompTIA N+ at Academy of Management and Science. At the same time he is working as a lead network and security engineer for PKI projects that have been authorized to be established by Bangladesh Government. A professional Information Security Management auditor according to ISO/IEC27001:2005. He has completed his graduation in Computer Security and Forensics from University of Bedfordshire, UK. And holding professional certification on CCNA and CEH while obtained training on VCP, CISSP, CHFI and ITILV3. Research interest goes to Information security and communication system improvement. 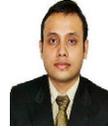

**Mohammad Iqbal** has vast experiences working as Telecommunication / LTE RF/ Netwok Engineer. He has completed MSC in Electrical Engineering withemphasis on Telecommunications from Blekinge Institute of Technology, Sweden in 2011. He holds professional certification on CCNA. He pursued his B.Sc. in Computer Science and Engineering from Northern University Bangladesh. 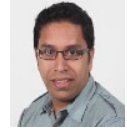